# Cloud Removal for Remote Sensing Imagery vai Spatial Attention Generative Adversarial Network

Heng Pan
*State Key Lab of CAD&CG, College of Computer Science*
*Zhejiang University*
Hangzhou, China
hengpan@zju.edu.cn

*Abstract*—Optical remote sensing imagery has been widely used in many fields due to its high resolution and stable geometric properties. However, remote sensing imagery is inevitably affected by climate, especially clouds. Removing the cloud in the high-resolution remote sensing satellite image is an indispensable pre-processing step before analyzing it. For the sake of large-scale training data，neural networks have been successful in many image processing tasks, but the use of neural networks to remove cloud in remote sensing imagery is still relatively small. We adopt generative adversarial network to solve this task and introduce the spatial attention mechanism into the remote sensing imagery cloud removal task, proposes a model named spatial attention generative adversarial network (SpA GAN), which imitates the human visual mechanism, and recognizes and focuses the cloud area with local-to-global spatial attention, thereby enhancing the information recovery of these areas and generating cloudless images with better quality. In the comparison experiment with the existing cloud removal models (conditional GAN, cycle GAN) on the open source RICE dataset, SpA GAN achieved the best performance on both peak signal to noise ratio (PSNR) and structural similarity index (SSIM). It proved that the spatial attention mechanism is effective for improving the quality of the cloud removal image and the superior performance of the model on the cloud removal task. The code of SpA GAN is https://github.com/Penn000/SpA-GAN_for_cloud_removal.

## I. INTRODUCTION

Optical remote sensing imagery has been widely used in many fields such as national defense security, environmental science, and weather monitoring due to its high resolution and stable geometric properties. However, when the remote sensing sensor carried by artificial satellite captures land information, it will inevitably be affected by the climate, especially clouds. Thus, removing the cloud in the high-resolution remote sensing imagery is an indispensable pre-processing step before analyzing it. The clouds to be removed from remote sensing imagery can be specifically classified into three categories, namely thin clouds, thick clouds, and cloud shadows, as shown in Fig. 1. The area covered by the thin clouds still keeps part of the ground feature captured by the remote sensing sensor, and the original information can be recovered from a single image. The information covered by the thick clouds is completely lost, which makes removing thick clouds from a single image become a condition-limited problem. Therefore, solving this problem often requires multitemporal data. Cloud shadows are usually caused by thick clouds blocking sunlight, they often appear with thick clouds at the same time. In this paper, we focus on removing thin clouds from a single optical remote sensing imagery as shown in Fig. 2.

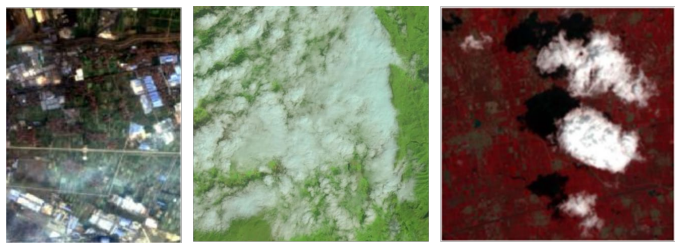

Fig. 1. Different type of clouds, from left to right is thin cloud, thick cloud, cloud shadow respectively.

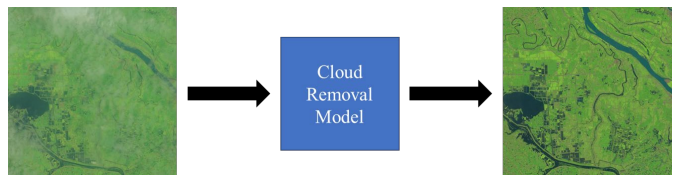

Fig. 2. Thin cloud removal for high-resolution remote sensing imagery, the cloud removal model takes a cloudy image as input and output a cloudless image.

The goal of cloud removal work is to recover the cloudless feature information from satellite images contaminated by clouds. Looking at this problem from a generalized perspective, it can also be understood as a kind of image denoising that clouds are the noise with regard to the surface objects. Recent years, relevant scholars have proposed their own solutions for this problem, including traditional image processing methods and deep learning methods. And in the field of computer vision, researches of removing fog, rain drop [17], watermark [18] and shade [19] have achieved a series of impressive results. These works have a use for reference to design the model that can remove clouds better.

Based on extensive investigation and previous work, we adopt generative adversarial network [1] to solve this task and introduce the spatial attention mechanism [2, 3] into the remote sensing imagery cloud removal field, and proposes a model named spatial attention generative adversarial networks or SpA GAN, which imitates the human visual mechanism, and recognizes and focuses the cloud area with local-to-global spatial attention, thereby enhancing the information recovery of these areas and generating cloudless images with better quality. In the comparison experiment with the existing cloud removal models on the open source RICE dataset [4], SpA GAN achieved the best performance on both peak signal to noise ratio (PSNR) [5] and structural similarity index (SSIM) [6]. It proved that the spatial attention mechanism is effective for improving the quality of the cloud removal image and the superior performance of the model on the cloud removal task.

## II. RELATED WORKS

### A. Cloud removal

The existing cloud removal methods can be divided into two categories: traditional image processing methods and deep learning methods. Traditional image processing methods hope to remove the cloud through pixel correction. Haze Optimized Transformation or HOT [7] is a classic thin cloud removal method which utilizes the high correlation between the blue and red bands in multispectral remote sensing imagery. However, HOT is sensitive to ground objects, easy to overcorrect and color distortion of RGB composite image. Improvement methods of HOT are then proposed [8]. Dark Channel Prior or DCP is also a classic thin cloud removal method which was first applied for image dehazing by [9] and achieved great success. DCP finds the prior knowledge of dark channel distribution through statistical analysis of the clear image library, and then uses it to infer the cloud model [10, 11]. Homomorphic filtering[12] is an image processing method that combines frequency filtering and grayscale transformation. It transforms the image into frequency domain via Fourier transform, and then uses a high-pass filter to filter the image and remove the thin clouds.

Traditional image processing methods mainly use the low-level features of the image, and the designed models often have limited performance. With the improvement of computing power, deep neural networks have made great progress in computer vision tasks, such as image restoration, image denoising and image super-resolution reconstruction. Some scholars try to use deep learning methods to solve the problem of high-resolution remote sensing imagery cloud removal. [13] applied the conditional generative adversarial networks [20] to the cloud removal field and proposed a model named multispectral conditional generative adversarial nets or McGANs. McGANs uses synthetic cloud satellite images and near-infrared band images to train the network so that it can automatically generate cloudless images from cloud images. In order to solve the trouble of the shortage of paired cloudless/cloudy data sets, [14] introduced the cycle generative adversarial network [15] and proposed a cloud removal model named cloud GAN. The core idea of cloud GAN is that model maps the cloudy image to the cloudless image through the generative network, and then maps it back to the cloudy image, the output should be similar to the original input. In addition, [16] proposes a cloud removal method by fusing synthetic aperture radar or SAR image data, which is a high-resolution imaging radar that can obtain high-resolution radar images similar to optical photography without being affected by climatic conditions.

### B. Generative adversarial networks

Generative Adversarial Networks [1] (GAN) is a deep learning model proposed by Goodfellow Ian at the 2014 Conference and Workshop on Neural Information Processing Systems (NeurIPS). GAN consists of two parts, the generator and the discriminator. The generator learns to obtain the data distribution of the target sample space, and the discriminator is used to evaluate the probability that a sample comes from the real data space instead of the generator generating data. Once proposed, GAN has attracted much attention. It is considered by the industry to be one of the most promising methods for learning complex data distribution. It currently plays an important role in many visual tasks like image generation, super-resolution reconstruction, data augmentation, and semantic segmentation. We believe the strong image generation power of GAN can transfer in cloud removal field.

## III. METHODOLOGY

GAN trains the generator and the discriminator in a game with each other. The generator is dedicated to generating data that makes the discriminator unable to distinguish whether the data comes from the training sample or the generator; while the discriminator is committed to learning to distinguish between true and false data. For generator $G$ and discriminator $D$, GAN can be defined as a minimax problem as:

$$\min_G \max_D V(D,G) = E_{x \sim p_{data}(x)}[\log(D(x))] \\ + E_{z \sim p_z(z)}[\log(1-D(G(z)))] \quad (1),$$

Here $x$ is real samples, $p_{data}$ is the distribution of $x$, $z$ is random noise, $p_z$ is the data distribution of $z$, $D(x)$ is the output of discriminator that represents the probability of $x$ is a real sample, $G(z)$ is the generated output of generator. Regard to cloud removal problem, we modify the primal GAN and propose the spatial attention GAN.

### A. Generator

The generative network of SpA GAN is a convolutional neural network called spatial attentive network (SPANet). Its overall structure is shown in Fig. 3. The input image first passes through a convolutional layer to extract the feature map, then passes through three standard residual blocks and four spatial attentive blocks (SAB), and then passes through two standard residual blocks and one convolutional layer, finally output the generated result which is the image with cloud removed. The SAB is shown in Fig. 3b, which includes three spatial attentive residual blocks (SARB) and one spatial attentive module (SAM) connected in parallel. The SAB module is used to discover and generate attention maps from the input feature maps. The attention map is a two-dimensional matrix, where the value of each element is a continuous value that indicates how much attention should be allocated to the pixel. The larger the value, the more attention should be given. It indicates the spatial distribution of the cloud which can guide the subsequent steps for cloud removal. The visualization of the attention map can be seen in Fig. 7. The SAM module is shown in Fig. 3d. It is a two-round, four-direction (up, down, left, right) recurrent neural networks with ReLU and identity matrix initialization (IRNN). The first round of IRNN is dedicated to generating a feature map that summarizes the contextual information of the location points from the input image; the second round of IRNN further collects non-local context information to generate a global perceptual feature map. The SARB module is shown in Fig. 3c. It removes clouds through negative residuals under the guidance of the attention map.

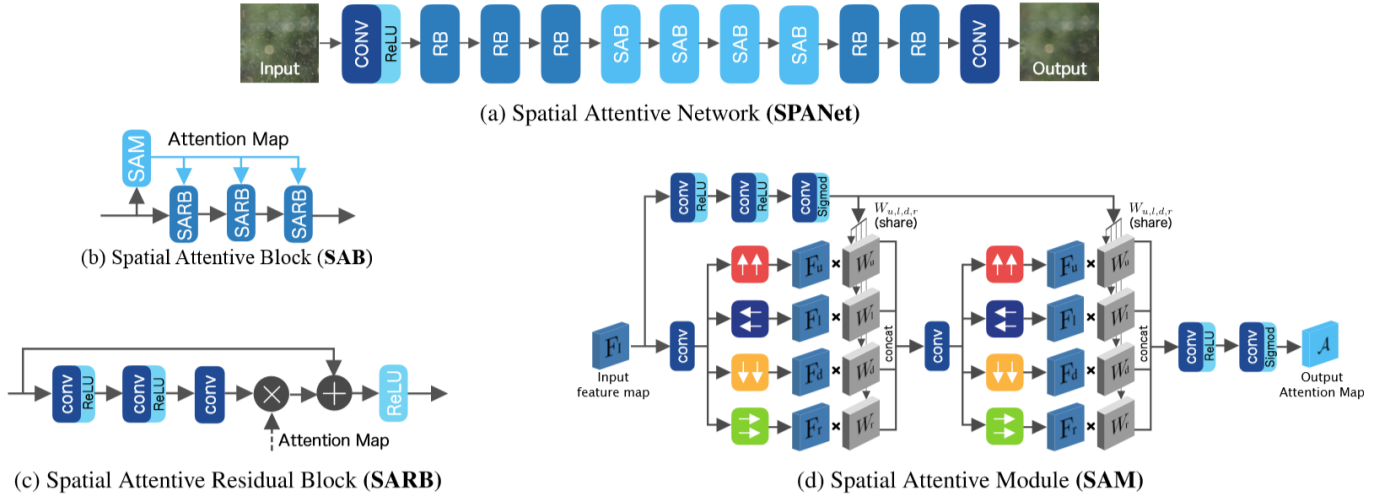

Fig. 3. Generator (a) of SpA GAN. It adopts three standard residual blocks to extract features, four spatial attentive blocks (SAB) (b) to identify cloud progressively in four stages, and two residual blocks to reconstruct a clean background. A SAB contains three spatial attentive residual blocks (SARBs) (c) and one spatial attentive module (SAM) (d).

### B. Discriminator

The discriminant network of SpA GAN is an ordinary convolutional neural network whose structure is shown in Fig. 4. Here $C$ represents the convolutional layer, $B$ represents the batch normalization layer, and $R$ represents the Leaky ReLU layer. The input of the network is a three-channel image, and the output is a flag of true or false, which means whether the input image is a real image or an image generated by the generator.

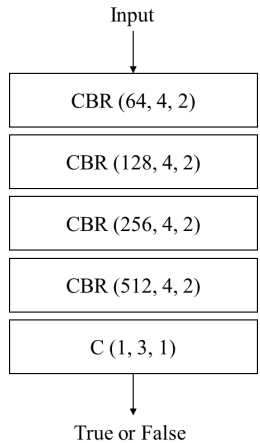

Fig. 4. The discriminator of SpA GAN.

### C. Loss

The total loss fo SpA GAN is:

$$L_{SpA} = \arg \min_G \max_D L_{cGAN}(G, D) + L_{L1}(G) + L_{Att} \quad (2)$$

The loss function consists of three parts where the first part is the loss function of conditional GAN, as shown in (3):

$$L_{cGAN}(G,D) = E_{x,y \sim p_{data}(x,y)}[\log D(x,y)] \\ + E_{x \sim p_{data}(x), z \sim p_z(z)}[\log(1 - D(x, G(x,z)))] \quad (3)$$

The second part is a standard L1 loss, which is used to measure the accuracy of each reconstructed pixel, as shown in (4), here $I_{in}$ is the input cloudy image, $I_{gt}$ is the ground truth, $\lambda_c$ is the weight of each channel contributes to the loss, which is set to 1 in our model. $G(I_{in})$ is the predicted result of generator, $C$, $H$, $W$ represent the number of channels, height and width of the image respectively.

$$L_{L1}(G) = \frac{1}{CHW} \sum_{c=1}^{C} \sum_{v=1}^{H} \sum_{u=1}^{W} \lambda_c \left| I_{gt}^{(u,v,c)} - G(I_{in})^{(u,v,c)} \right|_1 \quad (4)$$

The third part is attention loss, which is defined as (5). The matrix $A$ is the attention map generated by the spatial attentive module, and the matrix $M$ is the binary image of the cloud area, which is calculated by the difference between the cloudy and the cloudless image.

$$L_{Att} = \left\| A - M \right\|_2^2 \quad (5)$$

## D. Discuss

SpA GAN uses the spatial attention mechanism in the part of generative network. The attention map accumulates the global information of image during the generative process. Each location point will learn the information from its four-direction connected pixels, so do these pixels learn, spread in sequence, and finally learn the global information of the image. The attention map is not only used to guide the spatial attention residual block for cloud removal, its contribution is also reflected in the loss function which guides the training process of the model.

## IV. EXPERIMENTS

### A. Datasets

Cloud removal is an indispensable preprocessing step for high-resolution remote sensing imagery analysis, but deep learning methods are rarely used in the field of cloud removal. An important reason is the lack of data sets for training. Therefore, [4] provides a open source dataset named Remote sensing Image Cloud rEmoving (RICE) for cloud removal researching. The RICE dataset consists of two sub sets called RICE1 and RICE2, which is available on https://github.com/BUPTLdy/RICE_DATASET.

The RICE1 dataset contains 500 data samples with each sample having a cloudy image and a cloudless image under 512×512 resolution. The dataset is collected by Google Earth, and the cloudy/cloudless images are obtained by setting the cloud layer whether to display.

The RICE2 is constructed from Landsat 8 OLI/TIRS data by using LandsatLook images with georeferenced in Earth Explorer. LandsatLook images are full-resolution files derived from Landsat Level-1 data products. LandsatLook images include Natural Color Image, Thermal Image and Quality Image, here Natural Color Image and Quality Image are used in RICE2. [4] manually selected a cloudless image at the same location with a cloud image time less than 15 days apart to get the cloudless reference image. Finally, there are 736 groups of 512×512 images in the RICE2, and each group contains 1 cloudy, 1 cloudless, and 1 cloud mask image. The data samples of RICE are shown in Fig. 5.

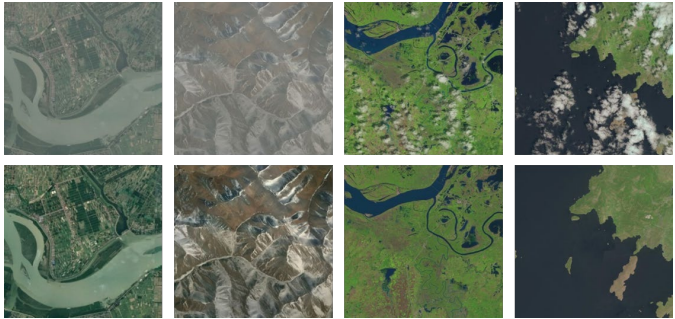

Fig. 5. The data samples of RICE, the first row is cloudy image, the second row is cloudless image. The first two columns of images belong to RICE1, the last two columns of images belong to RICE2.

### B. Evaluation metric

The input of model is a remote sensing image contaminate by cloud, and the output is a cloudless image after cloud removal. In order to measure the quality of the generated cloudless image and the cloud removal ability of the neural network model, peak signal to noise ratio (PSNR) [5] and structural similarity index (SSIM) [6] are widely used as image quality evaluation metrics.

*a) PSNR:* PSNR is the most widely and most common used objective measurement for evaluating image quality, whose calculation formula is:

$$PSNR = 10\log_{10}\left(\frac{(2^n-1)^2}{MSE}\right) \quad (6)$$

where $n$ is the bits of pixel value that $n$ is 8 for grayscale images. $MSE$ is the mean square error between the image $X$ and $Y$, calculated as:

$$MSE = \frac{1}{H \times W}\sum_{i=1}^{H}\sum_{j=1}^{W}(X(i,j)-Y(i,j))^2 \quad (7)$$

The value of PSNR generally situates in 20 to 40, the larger value represents the closer distance between predict image and ground truth image and the better prediction quality.

*b) SSIM:* SSIM is an evaluation metirc that measures the similarity of two images through three aspects: brightness, contrast, and structure, whose formula are (8), (9), (10) respectively:

$$l(X,Y) = \frac{2\mu_X\mu_Y + C_1}{\mu_X^2 + \mu_Y^2 + C_1} \quad (8)$$

$$c(X,Y) = \frac{2\sigma_X\sigma_Y + C_2}{\sigma_X^2 + \sigma_Y^2 + C_2} \quad (9)$$

$$s(X,Y) = \frac{\sigma_{XY} + C_3}{\sigma_X\sigma_Y + C_3} \quad (10)$$

Here $C_1, C_2, C_3$ are constants for the sake of avoiding divide zero error. $\mu, \sigma$ are the mean and variance of image, $\sigma_{XY}$ is the covariance of image $X$ and $Y$. Thus, the formula of SSIM is:

$$SSIM = l(X,Y) \cdot c(X,Y) \cdot s(X,Y) \quad (11)$$

The value range of SSIM is between 0 and 1, larger value means more similar between two images. If the value is 1, the two images are exactly same.

### C. Settings

For RICE dataset, we choose 400 images for training and 100 images for testing in RICE1, choose 588 images for training and 148 images for testing in RICE2, and set learning rate to 0.0004, minibatch to 1, epoch to 200 when training the model. In addition, we compare our SpA GAN with the existing cloud removal models including conditional GAN [20] and cycle GAN.

### D. Results and analysis

The results of conditional GAN, cycle GAN and SpA GAN on the RICE1 dataset are shown in Fig. 6. From left to right, each column represents cloudy image, conditional GAN generated result, cycle GAN generated result, SpA GAN generated result, and ground truth cloudless image respectively. Since almost all the clouds in RICE1 are thin clouds, the information of ground objects is not completely lost, every generated images after cloud removal retain the correct geometric structure and spatial information of the ground objects from visual perspective.

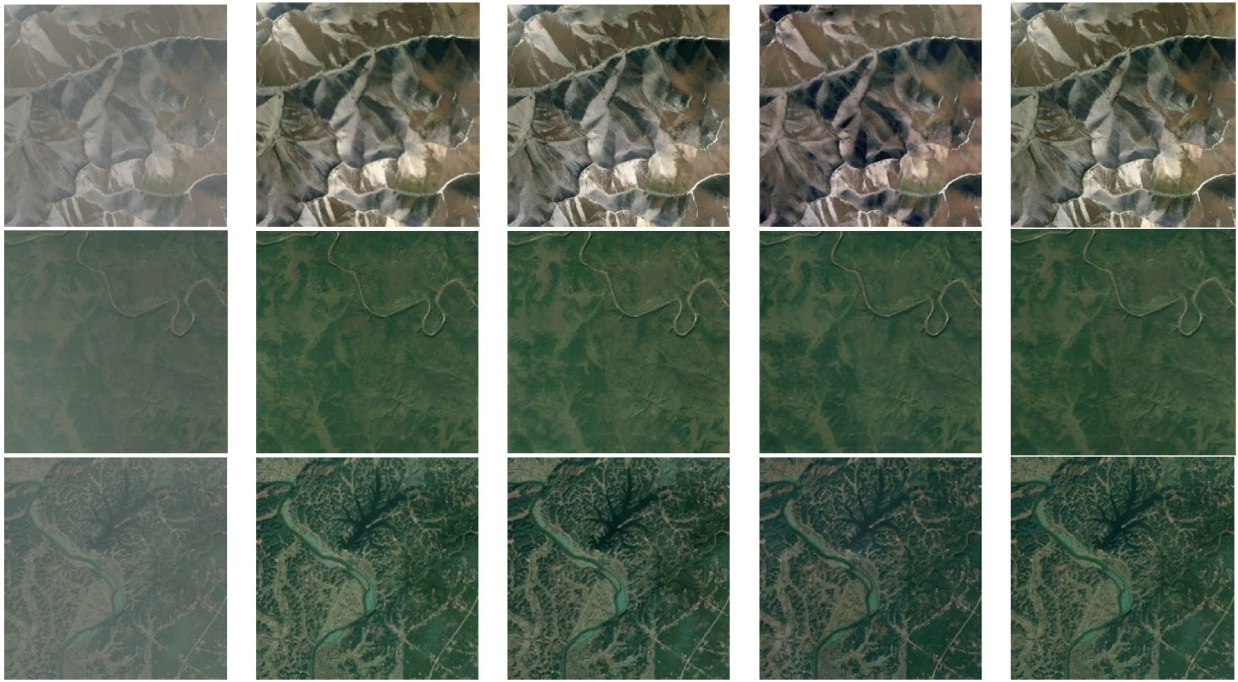

Fig. 6. The generated result on RICE1, from left to right, each column represents cloudy image, conditional GAN generated result, cycle GAN generated result, SpA GAN generated result, and ground truth cloudless image respectively.

The quantitative analysis results of conditional GAN, cycle GAN and SpA GAN on the test set are shown in Table 1. From Table 1, we can see that the PSNR and SSIM metrics of SpA GAN are 30.232dB and 0.954, respectively. Both have achieved the best performance and significantly outperform the other two models. It shows that the attention mechanism can observe cloud areas effectively and improve the performance of cloud removal. Cycle GAN has the lowest PSNR and SSIM values that demonstrates the necessity of paired discriminate information for high-resolution remote sensing imagery cloud removal.

TABLE I. QUANTITATIVE ANALYSIS ON RICE1 DATASET

| Model | Quantitative Metrics | |
|---|---|---|
| | *PNSR* | *SSIM* |
| Conditional GAN | 26.547 | 0.903 |
| Cycle GAN | 25.880 | 0.893 |
| SpA GAN | 30.232 | 0.954 |

The results of conditional GAN, cycle GAN and SpA GAN on the RICE2 dataset are shown in Fig. 8. From left to right, each column represents cloudy image, conditional GAN generative result, cycle GAN generative result, SpA GAN generative result, and ground truth cloudless image respectively. The RICE2 dataset contains a large number of images with thick clouds, where the ground objects information in the cloud-covered area is completely lost. The reconstruction of these pixels needs to learn from a large amount of similar data. As see in Fig. 8, cycle GAN removes the white area of thick cloud, but the corresponding ground objects is not well restored and image spatial continuity is disturbed. Although the image generated by conditional GAN ensures the spatial continuity, the predict cloudless image has fuzzy areas, and the details of the ground objects are not well restored. Compared with the aforementioned models, the cloudless image generated by SpA GAN retains more details and more consistency, which is visually closest to the ground truth. In addition, to the third row sample, we can found that the lake's shape and area between cloudy image and ground truth are inconsistent, but all three generative adversarial network models generate the correct lake that are consist with cloudy image. This shows the robustness and anti-noise ability of the generative adversarial network model on the task of cloud removal. On the other hand, it also shows the difficulty of obtaining cloudy/cloudless paired data at the same time and same place.

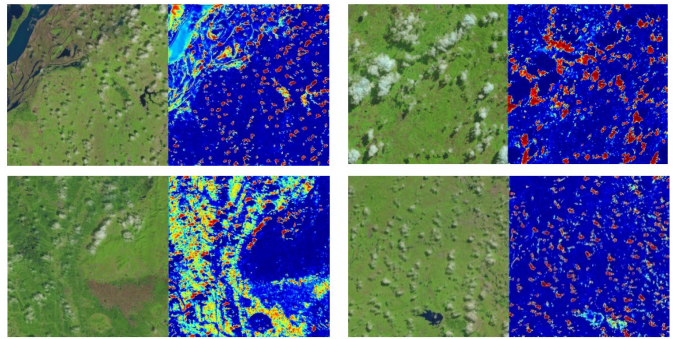

Fig. 7. The attention heatmap of SpA GAN.

The attention map generated by the SpA GAN during the cloud removal process is shown in Fig. 7. In the heatmap, the redder area means more attention are allocated, on the contrary the bluer area means less attention are allocated.

The quantitative analysis results of conditional GAN, cycle GAN and SpA GAN on the test set are shown in Table 2. The PSNR and SSIM of SpA GAN are 28.368dB and 0.906, respectively, both achieved the best performance during three models. PSNR increases 2.982dB compared to conditional GAN and increases 4.458dB compared to cycle GAN. SSIM increases 0.095 compared to conditional GAN and increases 0.113 compared to cycle GAN. The results illustrate the effectiveness of the attention mechanism in observing cloud areas and improving the performance of cloud removal once again.

TABLE II. QUANTITATIVE ANALYSIS ON RICE2 DATASET

| Model | Quantitative Metrics | |
|---|---|---|
| | *PNSR* | *SSIM* |
| Conditional GAN | 25.386 | 0.811 |
| Cycle GAN | 23.910 | 0.793 |
| SpA GAN | 28.368 | 0.906 |

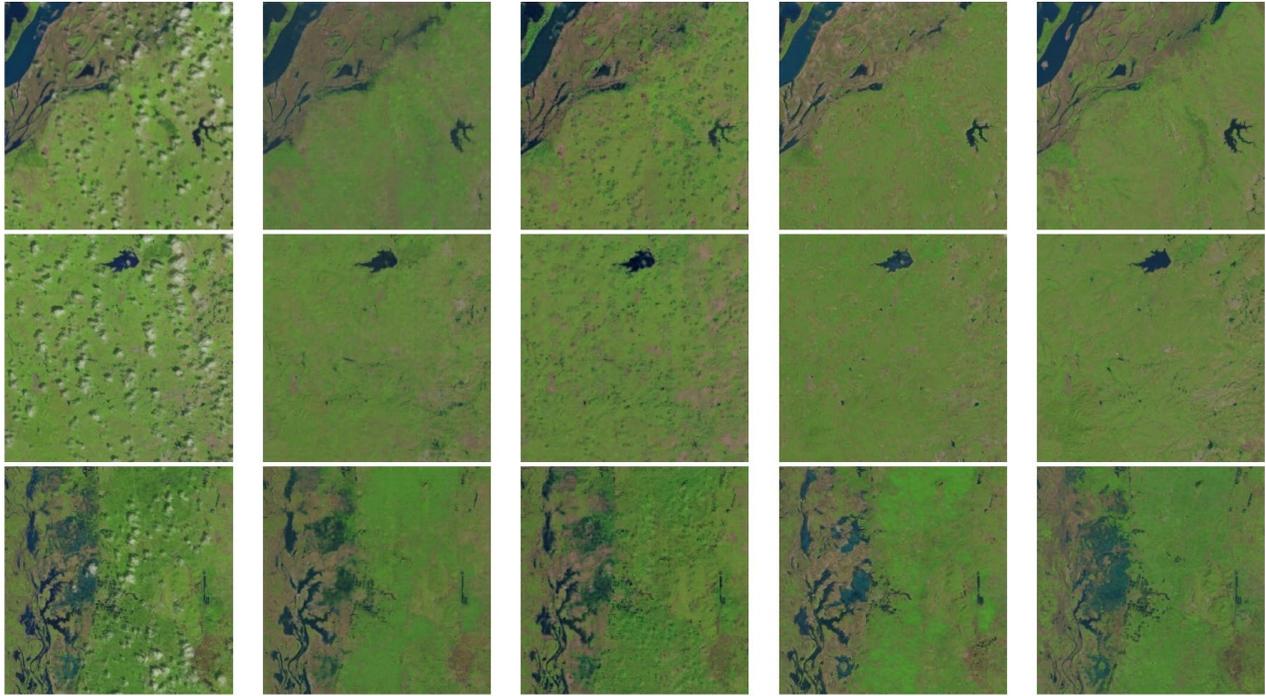

Fig. 8. The generated result on RICE2, from left to right, each column represents cloudy image, conditional GAN generated result, cycle GAN generated result, SpA GAN generated result, and ground truth cloudless image respectively.

## CONCLUSION

Remote sensing sensors are susceptible to interfered by climate, especially clouds when capture optical satellite images. It greatly reduces the availability of the satellite image data. Thus, cloud removal is a necessary preprocessing step before image analysis. Thanks to large-scale training data and powerful computing power, neural networks have been successful in many visual tasks, but the use of neural networks for remote sensing satellite cloud removal is still relatively small. This paper is the first that introduces spatial attention mechanism into the remote sensing imagery cloud removal task, and based on generative adversarial network, proposes the Spatial Attention Generative Adversarial Network or SpA GAN. Compare with conditional GAN and cycle GAN on the public RICE dataset, SpA GAN shows the best cloud removal ability that evaluated by PSNR and SSIM, which proves spatial attention mechanism is effective in improving the quality of cloud removal images.